\documentclass{article}

\usepackage{cite}
\usepackage[pdftex]{graphicx}
\graphicspath{{../pdf/}{../jpeg/}}
\DeclareGraphicsExtensions{.pdf,.jpeg,.png}
\usepackage{array}
\usepackage{amsmath}
\usepackage{amssymb}
\usepackage{mdwmath}
\usepackage{mdwtab}
\usepackage{eqparbox}
\usepackage{url}
\usepackage{hyperref}
\usepackage{multirow}
\usepackage{makecell}

\newenvironment{affiliations}{%
\setcounter{enumi}{1}%
\setlength{\parindent}{0in}%
\slshape\sloppy%
\begin{list}{\upshape$^{\arabic{enumi}}$}{%
    \usecounter{enumi}%
    \setlength{\leftmargin}{0in}%
    \setlength{\topsep}{0in}%
    \setlength{\labelsep}{0in}%
    \setlength{\labelwidth}{0in}%
    \setlength{\listparindent}{0in}%
    \setlength{\itemsep}{0ex}%
    \setlength{\parsep}{0in}%
    }
}{\end{list}\par\vspace{12pt}}

\title{The Multi-modality Cell Segmentation Challenge: Towards Universal Solutions}

\author{Jun Ma$^{1,2,3}$, Ronald Xie$^{1,3,4}$, Shamini Ayyadhury$^{5,6*}$, Cheng Ge$^{7*}$, \and 
Anubha Gupta$^{8*}$, Ritu Gupta$^{9*}$, Song Gu$^{10*}$, Yao Zhang$^{11*}$, Gihun Lee$^{12}$, \and
Joonkee Kim$^{12}$, Wei Lou$^{13,14}$, Haofeng Li$^{13}$, Eric Upschulte$^{15}$, Timo Dickscheid$^{15,16}$, \and
Jos\'e Guilherme de Almeida$^{17,18}$, Yixin Wang$^{19}$, Lin Han$^{20}$, Xin Yang$^{21}$, \and
Marco Labagnara$^{22}$, Vojislav Gligorovski$^{22}$, Maxime Scheder$^{22}$, \and
Sahand Jamal Rahi$^{22}$,  Carly Kempster$^{23}$, Alice Pollitt$^{23}$,  Leon Espinosa$^{24}$, \and 
T{\^a}m Mignot$^{24}$,  Jan Moritz Middeke$^{25,26}$, Jan-Niklas Eckardt$^{25,26}$, Wangkai Li$^{27}$, \and
Zhaoyang Li$^{28}$, Xiaochen Cai$^{29}$, Bizhe Bai$^{30}$,  Noah F. Greenwald$^{31}$,\and 
David Van Valen$^{32,33}$, Erin Weisbart$^{34}$, Beth A. Cimini$^{34}$, Trevor Cheung$^{1,35}$, \and
Oscar Br{\"u}ck$^{36,37}$, Gary D. Bader$^{4,5,6,38,39,40}$, and Bo Wang$^{1,2,3,38,41\dagger}$
}

\begin{document}

\maketitle

\begin{affiliations}
 \item Peter Munk Cardiac Centre, University Health Network, Toronto, ON, Canada
 \item Department of Laboratory Medicine and Pathobiology, University of Toronto, Toronto, ON, Canada
 \item Vector Institute, Toronto, ON, Canada
 \item Department of Molecular Genetics, University of Toronto, Toronto, ON, Canada
 \item Donnelly Centre, University of Toronto, Toronto, ON, Canada
 \item Princess Margaret Cancer Centre, University Health Network, Toronto, ON, Canada
 \item School of Medicine and Pharmacy, Ocean University of China, Qingdao, China
 \item Department of Electronics and Communications Engineering, Indraprastha Institute of Information Technology Delhi (IIITD), New Delhi, India
 \item Laboratory Oncology, Dr. BRA-IRCH, All India Institute of Medical Sciences, New Delhi, India
 \item Department of Image Reconstruction, Nanjing Anke Medical Technology Co., Ltd., Nanjing, China
 \item Shanghai Artificial Intelligence Laboratory, Shanghai, China
 \item Graduate School of AI, KAIST, Seoul, South Korea
 \item Shenzhen Research Institute of Big Data, Shenzhen, China
 \item Chinese University of Hongkong (Shenzhen), Shenzhen, China
 \item Institute of Neuroscience and Medicine (INM-1) and Helmholtz AI, Research Center J\"ulich, J\"ulich, Germany
 \item Faculty of Mathematics and Natural Sciences - Institute of Computer Science, Heinrich Heine University D\"usseldorf, D\"usseldorf, Germany
 \item European Molecular Biology Laboratory, European Bioinformatics Institute (EMBL-EBI), Hinxton, UK
 \item Champalimaud Foundation - Centre for the Unknown, Lisbon, Portugal
 \item Department of Bioengineering, Stanford University, Palo Alto, CA, USA
 \item Tandon School of Engineering, New York University, New York, NY, USA
 \item School of Biomedical Engineering, Health Science Center, Shenzhen University, Shenzhen, China
 \item Laboratory of the Physics of Biological Systems, Institute of Physics, \'Ecole Polytechnique F\'ed\'erale de Lausanne (EPFL), Lausanne, Switzerland
 \item School of Biological Sciences, University of Reading, Reading, UK
 \item Laboratoire de Chimie Bact\'erienne, CNRS–Universit\'e Aix-Marseille UMR, Institut de Microbiologie de la Méditerran\'ee, Marseille, France
 \item Department of Internal Medicine I, University Hospital Dresden, Technical University Dresden, Dresden, Germany
 \item Else Kroener Fresenius Center for Digital Health, Technical University Dresden, Dresden, Germany
 \item Department of Automation, University of Science and Technology of China, Hefei, China
 \item Institute of Advanced Technology, University of Science and Technology of China, Hefei, China
 \item Department of Computer Science and Technology, Nanjing University, Nanjing, China
 \item School of EECS, The University of Queensland, Brisbane, Australia
 \item School of Medicine, Stanford University, Palo Alto, CA, USA
 \item Division of Computing and Mathematical Science, Caltech, Pasadena, USA
 \item Howard Hughes Medical Institute, MD, USA
 \item Imaging Platform, Broad Institute of MIT and Harvard, Cambridge, MA, USA
 \item Department of Computer Science, University of Waterloo, Waterloo, Canada
 \item Hematoscope Lab, Comprehensive Cancer Center \& Center of Diagnostics, Helsinki University Hospital, Helsinki, Finland
 \item Department of Oncology, University of Helsinki, Helsinki, Finland
 \item Department of Computer Science, University of Toronto, Toronto, ON, Canada
 \item Lunenfeld-Tanenbaum Research Institute, Sinai Health System, Toronto, ON, Canada
 \item CIFAR Multiscale Human Program, CIFAR, Toronto, ON, Canada
 \item UHN AI Hub, University Health Network, Toronto, ON, Canada

$^*$These authors contributed equally. \\
$^\dagger$Corresponding author. Email: bowang@vectorinstitute.ai
\end{affiliations}

\maketitle

\newpage

\begin{abstract}
Cell segmentation is a critical step for quantitative single-cell analysis in microscopy images. Existing cell segmentation methods are often tailored to specific modalities or require manual interventions to specify hyper-parameters in different experimental settings. Here, we present a multi-modality cell segmentation benchmark, comprising over 1500 labeled images derived from more than 50 diverse biological experiments. The top participants developed a Transformer-based deep-learning algorithm that not only exceeds existing methods but can also be applied to diverse microscopy images across imaging platforms and tissue types without manual parameter adjustments. This benchmark and the improved algorithm offer promising avenues for more accurate and versatile cell analysis in microscopy imaging.
\end{abstract}

\section*{Introduction}

Cell segmentation is a fundamental task that is universally required for biological image analysis across a large number of different experimental settings and imaging modalities. 
For example, in multiplexed fluorescence image-based cancer microenvironment analysis, cell segmentation is the prerequisite for the identification of tumor sub-types, composition, and organization, which can lead to important biological insights~\cite{Nature2020Breast,Science22Lip,cell-colon-atlas}.  
However, the development of a universal and automatic cell segmentation technique continues to pose major challenges due to the extensive diversity observed in microscopy images. This diversity arises from variations in cell origins, microscopy types, staining techniques, and cell morphologies.
Recent advances ~\cite{hollandi2022nucReview} have successfully demonstrated the feasibility of automatic and precise cellular segmentation for specific microscopy image types and cell types, such as fluorescence and mass spectrometry images~\cite{Mesmer21-NatBio,lee2022cellseg},  differential interference contrast images of platelets~\cite{DIC-SciRep}, bacteria images~\cite{cutler2021omnipose} and yeast images~\cite{YeastMate,Yeaz}, but the selection of appropriate segmentation models remains a non-trivial task for non-expert users in conventional biology laboratories.

Efforts have been made towards the development of generalized cell segmentation algorithms~\cite{CellPose21-NM, cutler2021omnipose}. However, these algorithms were primarily trained using datasets consisting of gray-scale images and two-channel fluorescent images, lacking the necessary diversity to ensure robust generalization across a wide range of imaging modalities. For example, the segmentation models have struggled to perform effectively on RGB images, such as bone marrow aspirate slides stained with Jenner-Giemsa. Furthermore, these models often require manual selection of both the model type and the specific image channel to be segmented, posing challenges for biologists with limited computational expertise. 
In addition to directly training general cell segmentation models with large-scale labeled datasets, transfer learning-based algorithms are a complementary branch towards universal solutions, allowing biologists to rapidly train customized models on their own microscopy images. A prime example is Cellpose 2.0~\cite{stringer2022cellpose2}, which demonstrates the efficacy of adapting a pre-trained model to new images. Remarkably, it only requires 500–1,000 user-annotated image patches to achieve performance on par with models trained on thousands of image patches.

Biomedical image data science competitions have emerged as an effective way to accelerate the development of cutting-edge algorithms. Several successful competitions have been specifically organized for microscopy image analysis, such as the cell tracking challenge (CTC)~\cite{celltrack-nm17,celltrack-nm23}, the Data Science Bowl (DSB) challenge~\cite{NM-DSB18}, and Colon Nuclei Identification and Counting Challenge (CoNIC)~\cite{CoNIC}. These competitions have played a crucial role in expediting the adoption of modern machine learning and deep learning algorithms in biomedical image analysis. However, it is worth noting that these challenges have primarily focused on a limited subset of microscopy image types. 
For example, the CTC primarily concentrated on label-free images, thereby excluding stained images such as multiplexed immunofluorescent images. Similarly, the DSB challenge emphasized nucleus segmentation in fluorescent and histology images while disregarding phase-contrast and differential interference contrast images. The segmentation task in the CoNIC challenge is also limited to nucleus segmentation in H\&E stained images.
Consequently, the algorithms developed through these competitions are often tailored to handle only specific types of microscopy images, limiting their generalizability. 
Moreover, the evaluation metrics used in these challenges predominantly prioritize segmentation accuracy, while neglecting algorithm efficiency. As a result, the pursuit of higher accuracy scores often leads to the adoption of computationally demanding approaches. For instance, the CTC top-performing algorithms~\cite{celltrack-nm23} employed customized models for each dataset in the cell segmentation task, while the DSB winning algorithm~\cite{NM-DSB18} used an ensemble of 32 models. Such resource-intensive strategies hinder the wide deployment of these algorithms in biology practice.

% P4. Our contributions
To overcome the aforementioned limitations and foster the development of universal and efficient cell segmentation methods for microscopy images, we took the initiative to organize a global challenge at the Conference on Neural Information Processing Systems (NeurIPS). As one of the largest international conferences in the field of artificial intelligence (AI), NeurIPS provided an ideal platform for this endeavor.
Participants in the challenge were provided with a diverse training set and a separate tuning set to develop and refine their cell segmentation algorithms. During the testing phase, participants were required to package their algorithms as Docker containers, enabling the challenge organizers to evaluate them on a carefully curated holdout testing set on the same computing platform.
Importantly, the holdout testing set incorporated images from new biological experiments, aiming to assess the algorithms' ability to generalize effectively to previously unseen data. Additionally, the testing set included two whole-slide images, serving as a means to evaluate the algorithms' suitability for handling large-scale images.
Different from existing challenges that focused on specific microscopy image types, this initiative represents the first instance where cell segmentation algorithms were challenged to efficiently handle a broad spectrum of microscopy images with one single model and generalize to new images without manual intervention. 

% figure 1
\begin{figure*}[!htbp]%
\centering
\includegraphics[scale=0.18]{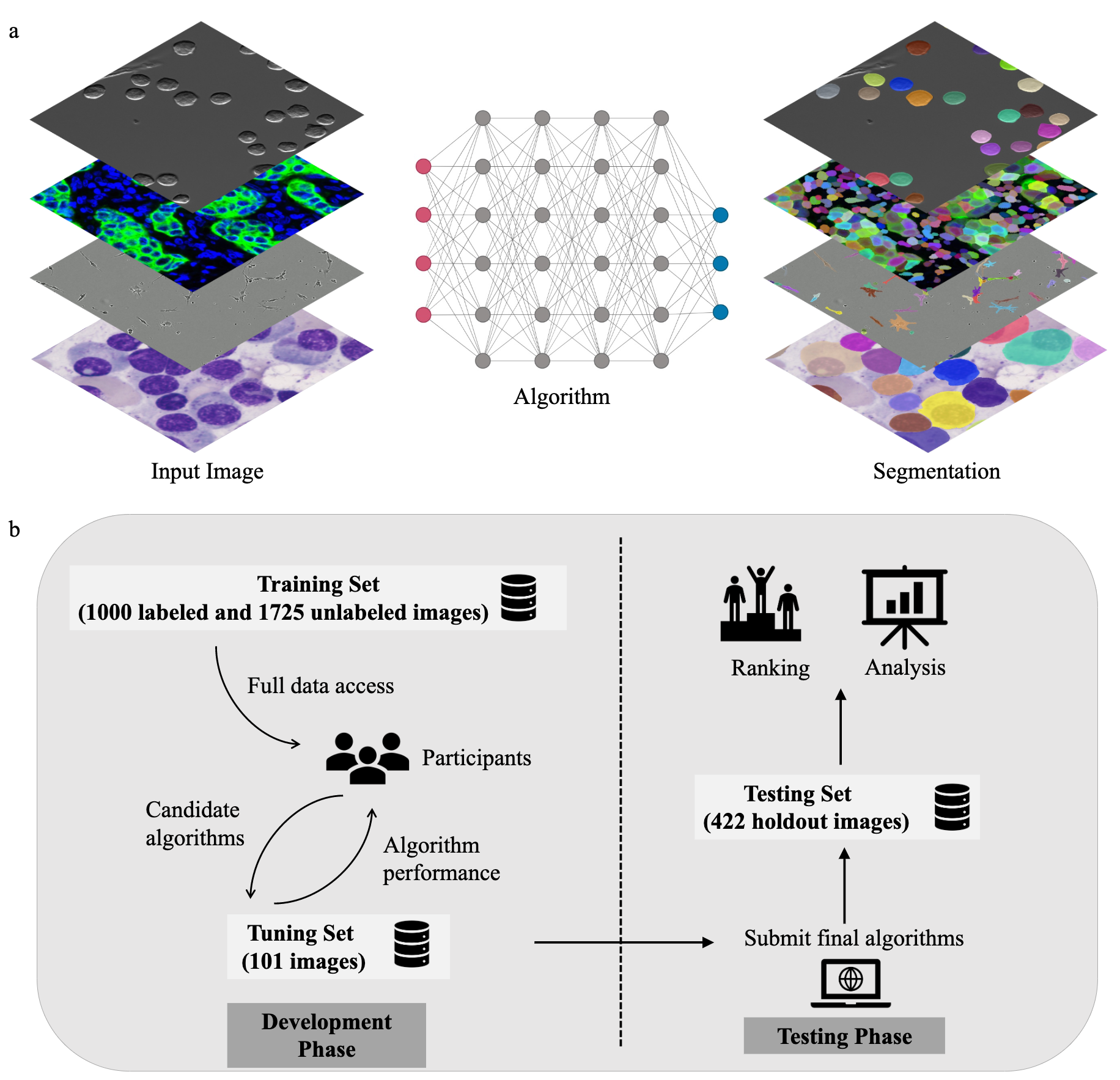}
\caption{\textbf{Overview of the challenge task and pipeline.} \textbf{a,} The challenge aims to facilitate the development of universal cell segmentation algorithms that can segment a wide range of microscopy images without manual intervention. \textbf{b,} The challenge contains two phases. During the development phase, participants develop automatic segmentation algorithms based on 1000 labeled images and 1725 unlabeled images. The algorithms can be evaluated on a tuning set with 101 images and the online evaluation platform will automatically return back the quantitative performance. During the testing phase, each team can submit one algorithm via the Docker container as the final solution, which is independently evaluated on the holdout testing set with 422 images to obtain ranking results.
}\label{fig:challenge}
\end{figure*}

\section*{Results}
\subsection*{Challenge design: towards universal and efficient cell segmentation algorithms}
The primary objective of this challenge was to benchmark universal algorithms capable of accurately segmenting cells from a wide range of microscopy images obtained from various imaging platforms and tissue types, without requiring additional parameter tuning (Fig.~\ref{fig:challenge}a). The algorithms were expected to operate in a fully automatic manner, generating cell instance masks where each cell is assigned a unique label. 
The challenge comprised two phases (Fig.~\ref{fig:challenge}b). In the development phase, participants were provided with a dataset consisting of 1000 microscopy images, each accompanied by annotated cell masks. Recognizing the potential benefits of leveraging unlabeled data to enhance model performance~\cite{imperfect-data-MIA20}, we also made an additional set of 1725 unlabeled images available for participants to utilize.
Participants were given the flexibility to decide whether to incorporate this unlabeled dataset into their algorithms. This setup aligns with real-world scenarios encountered in biological research, where only a limited number of labeled images are typically available alongside a wealth of unlabeled images.

To facilitate timely model validation, a separate tuning set containing 101 images was provided to participants, but the corresponding annotations were not disclosed. Instead, we established an online evaluation platform, enabling participants to upload their segmentation results and receive evaluation scores. These scores were made publicly available on a leaderboard, enabling direct comparisons among participants and their algorithms throughout the development phase.

In the subsequent testing phase, the top 30 teams, as ranked on the public tuning set leaderboard, were invited to make the testing submission. 
The testing set remained hidden from participants, aiming to avoid potential label leaking and cheating. 
To ensure standardized evaluation, participants were required to package and submit their algorithms as Docker containers.
Challenge organizers run the submitted Docker containers on the holdout testing set comprising 422 microscopy images. Out of the 30 invited top teams, 28 teams made successful submissions, while one team did not submit and another team submitted after the deadline, making their submission ineligible for final ranking.
To ensure a fair comparison, we executed the Docker containers sequentially on the same workstation. The running time for each image was recorded, alongside the corresponding segmentation accuracy score. Both of them were used for the final ranking and subsequent analysis of the algorithms (Methods).

% figure 2
\begin{figure*}[!htbp]%
\centering
\includegraphics[scale=0.11]{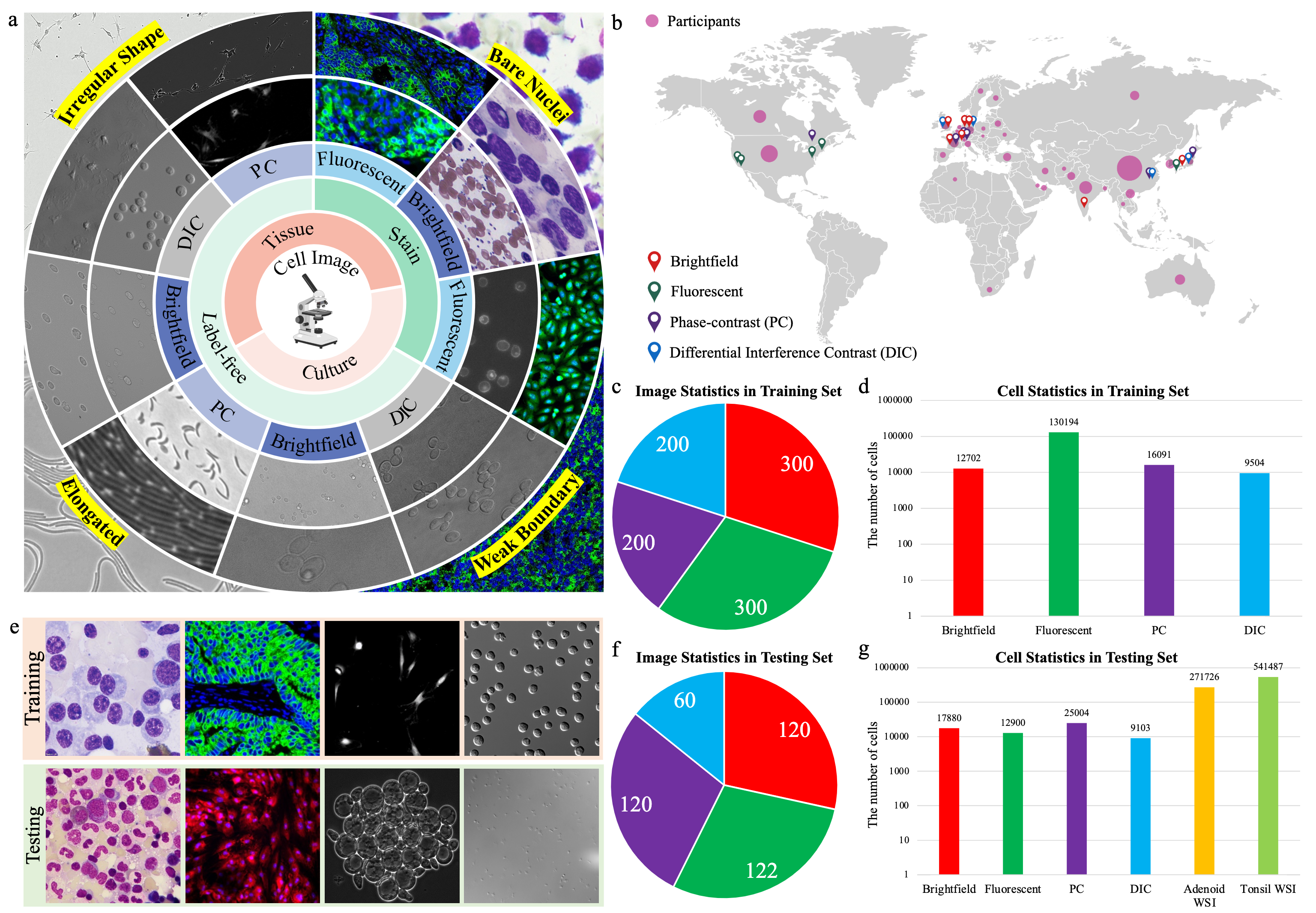}
\caption{\textbf{Dataset overview.} \textbf{a,} The challenge provides a diverse microscopy image dataset that includes tissue cells, cultured cells, label-free cells, stained cells, and different microscopes (i.e., brightfield, fluorescent, phase-contrast (PC), and (Differential Interference Contrast) DIC). \textbf{b,} The geographical distribution of data sources and challenge participants. The red, green, purple, and blue address icons denote the countries or regions where the brightfield, fluorescent, phase-contrast, and differential interference contrast image datasets are from, respectively. The size of the pink circle in each country is proportional to the number of participants from the corresponding country. \textbf{c,} The number of images in the training set. \textbf{d,} The number of labeled cells in the training set. \textbf{e}, Randomly selected examples (from left to right: brightfield, fluorescent, PC, and DIC images) from the training set (the 1st row) and testing set (the 2nd row). \textbf{f,} The number of images in the testing set. \textbf{g,} The number of cells in the testing set. There are two fluorescent whole-slide images (WSI) in the testing set. 
}\label{fig:dataset}
\end{figure*}

\subsection*{Challenge data: a large and diverse multi-modality microscopy image dataset}
Data diversity plays a pivotal role in constructing generalist microscopy image segmentation models~\cite{ma2023NM-SegCom}. In this challenge, we incorporated the diversity of microscopy images from four dimensions: cell origins, staining methods, microscope types, and cell morphologies (Fig.~\ref{fig:dataset}a). 
First, the origin of cells in microscopy images varies substantially, as they can derive from diverse tissues or exist within cell cultures under various conditions. This introduces considerable variability, as cells within tissues tend to be densely packed and spatially organized, whereas cells in culture are often sparsely distributed and randomly positioned.
Second, the choice of staining methods, such as Jenner-Giemsa in brightfield microscopy or the utilization of specific antibodies in fluorescent microscopy, further contributes to the diversity by highlighting different cellular structures or proteins.
Third, the use of different microscope types, such as brightfield, fluorescent, phase-contrast (PC), and differential interference contrast (DIC), introduces substantial differences in image characteristics, textures, and associated artifacts. 
Fourth, cell morphologies exhibit substantial variations across different cell types. While most cells tend to have a round shape, certain cells may display elongated or irregular shapes.

We curated a diverse microscopy image dataset by collecting images (and annotation if available) from over 20 biology laboratories, including more than 50 different biological experiments (Supplementary Table 1-3). This comprehensive dataset encompassed four common microscopy image modalities: brightfield, fluorescent, phase-contrast (PC), and differential interference contrast (DIC).
The challenge garnered a large number of interest and participation, attracting over 400 participants from 37 different countries, reflecting the global reach and impact of the challenge (Fig.~\ref{fig:dataset}b).
The training set contained a total of 1000 images, with 300 images each in the brightfield and fluorescent categories, and 200 images each in the PC and DIC categories (Fig.~\ref{fig:dataset}c). The annotated dataset contained 12,702 cells in brightfield images, 130,194 cells in fluorescent images, 9,504 cells in PC images, and 16,091 cells in DIC images (Fig.~\ref{fig:dataset}d). Notably, the higher cell count in fluorescent images compared to other modalities can be attributed to the denser distribution of cells observed in the collected fluorescent images.

Fig.~\ref{fig:dataset}e shows four microscopy images randomly selected from each modality in the training set and testing set. In order to assess the algorithm's generalization capabilities, all testing images were sourced from new biological experiments, including some that featured previously unseen tissues or cell types not present in the training set. The testing set consisted of 120 brightfield images, 122 fluorescent images, 120 phase-contrast (PC) images, and 60 differential interference contrast (DIC) images (Fig.~\ref{fig:dataset}f). These quantities were determined based on the available images collected for the challenge. 
The number of cells in the testing set was comparable to or greater than that of the training set (Fig.~\ref{fig:dataset}g). Additionally, the fluorescent image subset of the testing set included two whole-slide images (WSIs), which served the purpose of evaluating the algorithms' ability in handling large-scale imaging datasets.

In comparison to previous datasets utilized in cell segmentation challenges~\cite{segpc-mia, celltrack-nm23} and nucleus segmentation challenges~\cite{NM-DSB18,CoNIC}, our dataset exhibits substantially enhanced diversity and encompasses a larger number of labeled cells. This extensive dataset serves as a fertile ground for fostering the development of advanced cell segmentation algorithms, enabling researchers to explore and innovate in the field.

\subsection*{Algorithm overview: the Transformer-based algorithm achieved superior performance}
All algorithms in this challenge employed deep learning-based approaches, a prevailing trend considering the remarkable performance achieved in various specific cell segmentation tasks~\cite{Mesmer21-NatBio, Yeaz, celltrack-nm23}, as well as in recent generalist cell segmentation algorithms~\cite{stringer2022cellpose2, cutler2021omnipose}. Existing algorithms predominantly relied on Convolutional Neural Networks (CNNs) such as U-Net~\cite{UNet19-NM} and DeepLab~\cite{chen2017deeplab}. However, it is worth noting that these CNN-based cell segmentation models exhibited limited generalization capability when confronted with the task of segmenting diverse images without additional human intervention, such as manual selection of channels or model fine-tuning, as demonstrated in the following sections.

In contrast, Transformers~\cite{attention-nips17}, a new type of deep learning network integrating attention mechanisms for feature extraction, have exhibited robust performance and generalization capabilities across various computer vision tasks~\cite{ViT2020,MedSAM}. However, the potential of Transformers in biological image analysis remains relatively unexplored~\cite{ma2023NM-SegCom}.
Distinguished from existing benchmarks~\cite{NM-DSB18, celltrack-nm23}, our challenge provided a significantly larger and more diverse microscopy image dataset. Leveraging this unprecedented dataset and a meticulously designed benchmark, Transformer-based deep learning models emerged as exceptional algorithms and achieved notably superior performance.

\textbf{Best-performing algorithm.}
Lee et al.~\cite{top1} (T1-osilab) proposed a Transformer-based framework to harmonize model-centric and data-centric approaches. The model architecture used SegFormer~\cite{xie2021segformer} and multiscale attention network~\cite{MANet} as the encoder and decoder. The SegFormer encoder was a hierarchical Transformer, enabling the extraction of both coarse and fine-grained features. The decoder contained position-wise attention blocks and multiscale fusion attention blocks for feature map fusion. The model output comprised two separate heads for cell recognition and distinction, originally proposed in~\cite{CellPose21-NM}. 
The model underwent a two-step training process. It was first pre-trained on public microscopy images and then fine-tuned on the challenge dataset with cell-aware data augmentation. Additionally, cell memory reply~\cite{memory-replay}, concatenating the images from the pre-training and fine-tuning datasets in each mini-batch, was used to avoid catastrophic forgetting during fine-tuning (Methods).

\textbf{Second-best-performing algorithm.}
Lou et al.~\cite{top2} (T2-sribdmed) first divided the images into four distinct categories based on low-level image features (e.g., intensities) in an unsupervised way. Then, class-wise cell segmentation models were trained for each category. The model employed U-Net-like architecture where ConvNeXT~\cite{ConvNext} was used as the building blocks. To address the diverse cell morphologies, two distinct decoder heads were employed. One decoder predicted the cell distance map and semantic map, effectively segmenting round-shaped cells, while the other decoder predicted the cell gradient map to handle cells with irregular shapes. The training process involved pre-training the model on the entire dataset, followed by fine-tuning on each of the four categories, resulting in the creation of four models. During inference, the image was initially classified into one of the four categories, and subsequently, the corresponding model was used to perform the segmentation process (Methods).

\textbf{Third-best-performing algorithm.}
Upschulte et al.~\cite{top3} (T3-cells) designed an uncertainty-aware contour proposal network, employing ResNeXt-101~\cite{ResNeXt} to extract multiscale features from images, which were then processed through four decoder heads. A classification head identified potential cell locations, while a contour regression head predicted sparse cell contours. To further improve accuracy, a refinement regression head was employed to revise the pixels within the cell contour. In addition, they incorporated an uncertainty head to estimate prediction confidence, which played a crucial role in the non-maximum suppression post-processing. This incorporation of uncertainty information effectively facilitated the removal of redundant contour proposals and enhanced segmentation accuracy (Methods). 

% tables
\begin{table}[htb]
\caption{Characteristics of top three best-performing algorithms in pre-processing, data augmentation, network architecture, and post-processing. Abbreviation: Intensity Normalization (IN), Patch Sampling (PS), Intensity and Spatial data augmentation (IS), External Datasets (ED), Unlabeled Data (UD), Non-Maximum Suppression (NMS), Test-time Augmentation (TTA). `-' denotes it is not used.}
\label{tab:methods}
\centering
\resizebox{\textwidth}{!}{
\begin{tabular}{c|cc|cccl|ll|l}
\hline
\multirow{2}{*}{Team} & \multicolumn{2}{c|}{Pre-processing} & \multicolumn{4}{c|}{Data Augmentation}                                                                                                                                & \multicolumn{2}{c|}{Network Architecture}                                                                                                                                                                             & \multicolumn{1}{c}{\multirow{2}{*}{Post-processing}}                                                                                  \\ \cline{2-9}
                      & IN              & PS             & IS                                                                                                          & ED & UD  & \multicolumn{1}{c|}{Others}                   & \multicolumn{1}{c}{Encoder Backbone} & \multicolumn{1}{c|}{Decoder Heads}                                                                                                                                                         & \multicolumn{1}{c}{}                                                                                                                  \\ \hline
\begin{tabular}[c]{@{}c@{}}T1 \\ osilab \\ \cite{top1}\end{tabular}                    & \checkmark                  & \checkmark              & \checkmark   & \checkmark  & \multicolumn{1}{l}{-}     & \begin{tabular}[c]{@{}l@{}}Cell-wise intensity \\ perturbation; \\ Boundary exclusion; \\ Oversample \\ minor modality;\end{tabular}    & SegFormer              & \begin{tabular}[c]{@{}l@{}}Cell probability head;\\ Gradient fields \\ regression head;\end{tabular}                                                                    & \begin{tabular}[c]{@{}l@{}}Gradient tracking; \\ Exclud small cells;\\ Fill holes;\\ TTA;\end{tabular}                      \\ \hline
\begin{tabular}[c]{@{}c@{}}T2 \\ sribdmed \\ \cite{top2}\end{tabular}                     & \checkmark                  & \checkmark              & \checkmark                                                                                                                                    & \checkmark  & \checkmark    & -                  & ConvNeXt               & \begin{tabular}[c]{@{}l@{}}Cell probability head; \\ Radial distance head; \\ Gradient fields \\ regression head;\end{tabular}                                          & \begin{tabular}[c]{@{}l@{}}NMS; \\ Watershed;\end{tabular}                                                                            \\ \hline
\begin{tabular}[c]{@{}c@{}} T3 \\ cells \\ \cite{top3}  \end{tabular}                  & \checkmark                  & \checkmark              & \checkmark                                                                  & \checkmark  & \checkmark        & \begin{tabular}[c]{@{}l@{}}Cell-aware rescaling\end{tabular}              & ResNeXt-101            & \begin{tabular}[c]{@{}l@{}}Classification head; \\ Contour \\ regression head; \\ Local refinement \\ regression head; \\ Boundary uncertainty \\ estimation head;\end{tabular} & \begin{tabular}[c]{@{}l@{}}NMS; \\ Convert contours \\ to masks; \\ Region growing;\end{tabular} \\ \hline
\end{tabular}
}
\end{table}

\textbf{Other strategies.} Table~\ref{tab:methods} summarizes the strategies employed by the top three teams. Given the considerable variation in image intensity and size across different modalities, all these teams adopted intensity normalization techniques (e.g., scaling the intensity to [0, 255]) during pre-processing and opted for patch-based sampling for model training. To enhance the model generalization ability and mitigate the risk of overfitting, diverse data augmentation methods were utilized. In addition to using external datasets, teams T2 and T3 leveraged the unlabeled data for model pre-training. Despite the common adoption of an encoder-decoder framework to construct networks, the top teams showcased variations in their choice of backbone networks and decoder heads. Consequently, the corresponding post-processing methods exhibited diversity.

% figure 3
\begin{figure*}[!htbp]%
\centering
\includegraphics[scale=0.3]{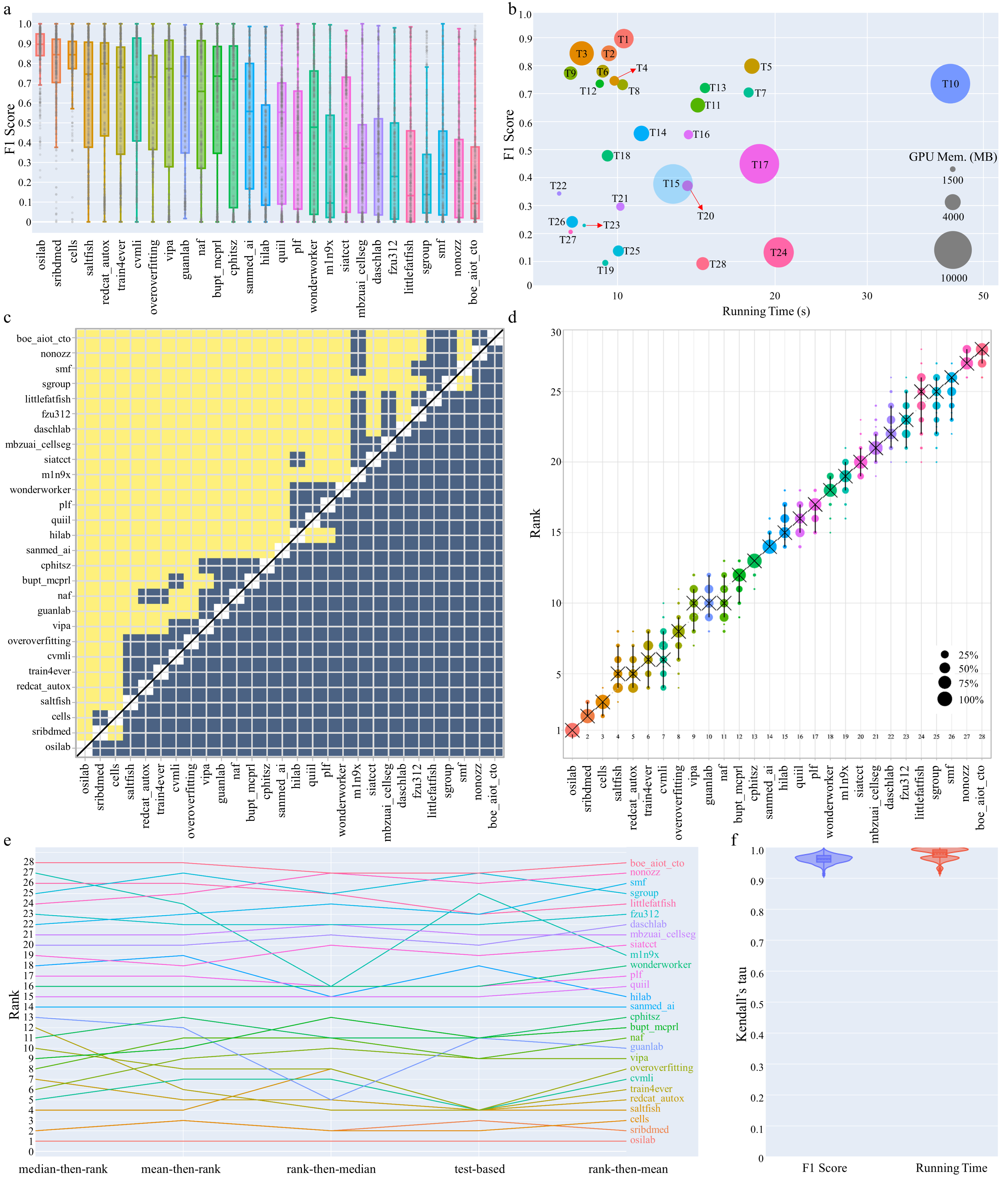}
\caption{\textbf{Evaluation results of 28 algorithms on the holdout testing set.} \textbf{a,} Dot and box plot of the F1 scores on the testing set (n=422 independent images). The box plots display descriptive statistics across all testing cases, with the median value represented by the horizontal line within the box, the lower and upper quartiles delineating the borders of the box, and the vertical black lines indicating the 1.5 interquartile range. \textbf{b,} The top algorithms achieve a good trade-off between segmentation accuracy (y-axis) and efficiency (x-axis). The circle size is proportional to GPU memory consumption. 
\textbf{c,} Pairwise significant test results (one-sided Wilcoxon signed rank test) show that the winning algorithm is significantly better than the other algorithms. 
\textbf{d,} Blob plot for visualizing ranking stability based on bootstrap sampling. The median area of each blob is proportional to the relative frequency of achieved ranks across 1000 bootstrap samples. The median rank for each algorithm is indicated by a black cross. 95\% bootstrap intervals across bootstrap samples are indicated by black lines. \textbf{e,} The winning algorithm holds the first place across five different ranking schemes. \textbf{f,} The high Kendall's tau scores indicate that the ranking results are stable. The volin plot shows descriptive statistics with the median value represented by the horizontal solid line within the box, the mean value represented by the horizontal dashed line the lower and upper quartiles delineating the borders of the box, and the vertical black lines indicating the 1.5 interquartile range.
}\label{fig:eval}
\end{figure*}

Next, we present the quantitative results of the 28 algorithms on the holdout testing set.
Fig.~\ref{fig:eval}a (Supplementary Table 4) shows a comparative view of F1 scores across 28 algorithms on the testing set. The scores are presented in the form of a dot and box plot, offering insights into both their central tendency and dispersion of the scores.
The top three algorithms surpass other algorithms by a clear margin, resulting in median F1 scores of 89.7\% (interquartile range (IQR): 84.1-94.8\%), 84.5\% (IQR: 70.6-92.3\%), and 84.4\% (IQR: 77.4-91.1\%), respectively. Of particular note is the performance of the winning algorithm (T1-osilab). It stands apart not merely for its superior median F1 score, but also for the reduced number of outliers in its score distribution, suggesting a heightened level of robustness in its performance.

The bubble plot (Fig.~\ref{fig:eval}b) presents the median F1 score, running time, and the maximum GPU memory consumption of 28 algorithms, which can provide insights into the trade-off between algorithm accuracy and efficiency. Most algorithms optimized the efficiency, enabling them to finish the inference within 13s. It is essential to mention that this time metric also included the Docker starting time, hence the actual inference time is considerably shorter. For instance, the best-performing algorithm (T1-osilab) achieved an inference time of approximately 2 seconds for an image size of $1000\times1000$. Additionally, the median maximum GPU memory consumption was 3099MB (approximately \$500), suggesting that these algorithms are affordable for practical deployment. This favorable combination of accuracy and efficiency makes them well-suited for real-world applications in biological image analysis.

We also performed a statistical significance analysis for the 28 algorithms (Fig.~\ref{fig:eval}c). Each team was compared with the other teams based on the one-sided Wilcoxon signed rank test. Yellow shading indicates that the F1 scores of the algorithm on the x-axis are significantly superior ($p<0.05$) to those from the algorithm on the y-axis, while blue shading indicates no significant difference between the two algorithms. The winning algorithm is significantly better than all the others. The 2nd algorithm and the 3rd algorithm obtain comparable performances with no significant differences, but they are significantly superior to other teams. 

Furthermore, we analyzed the ranking stability based on bootstrap sampling (1000 times). The results are visualized by blob plot (Fig.~\ref{fig:eval}d). The blob area is proportional to the relative frequency of achieved ranks across the bootstrap samples and the median rank for each algorithm is indicated by a black cross. The winning algorithm has a blob area of 100\%, indicating that it outperforms other algorithms in all the bootstrap samples.  The 2nd and 3rd best-performing algorithms still obtain better rank than other algorithms with a clear gap while the 2nd best-performing algorithm has a lower median rank than the 3rd best-performing algorithm. Moreover, we compared the ranks of the 28 algorithms based on different ranking schemes (Fig.~\ref{fig:eval}e): median-then-rank, mean-then-rank, rank-then-median, statistical significance test-based ranking, and rank-then-mean (Methods). The winning algorithm consistently ranks first place across all the ranking schemes, while most other teams have fluctuations in terms of the rank.

Finally, we analyzed the ranking stability of the employed metrics. The ranking list based on the full testing set is pairwise compared with the ranking lists based on the
individual sample in the 1000 bootstrap samples. Kendall's tau correlation is computed as a quantitative metric (Fig.~\ref{fig:eval}f, Extended Data Fig. 2). It can be found that Kendall's $\tau$ scores are very close to 1 for both F1 scores and running time, indicating a high degree of ranking agreement. Additionally, the compact distributions of these scores further confirm the stability of the ranking results with respect to sampling variability. These findings provide robust evidence that the obtained rankings are highly consistent and reliable across different samples.

%%%%%%%%%%%%%%%%%%%%%%%%%%%%%%%%%%%%%%%%%% PK

% figure 4
\begin{figure*}[!htbp]%
\centering
\includegraphics[scale=0.16]{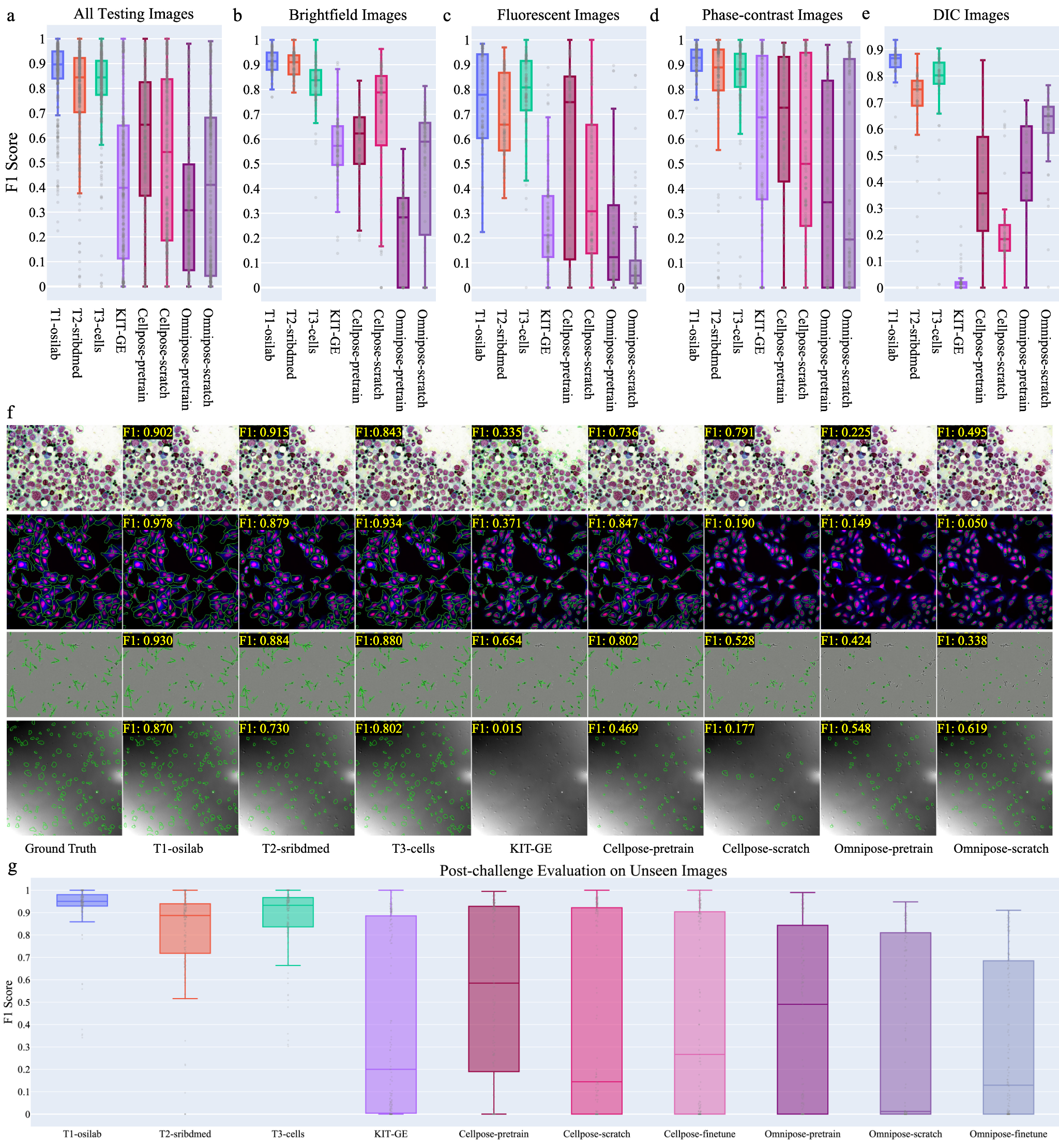}
\caption{\textbf{Quantitative and qualitative comparison between the top three algorithms and state-of-the-art generalist cell segmentation algorithms: KIT-GE (top solution in the segmentation benchmark of the cell tracking challenge), Cellpose, Omnipose, and their variants under different training strategies.} Dot and box plot of the F1 scores on the \textbf{a,} whole testing test (n=422 independent images);  \textbf{b,} brightfield images (n=120 independent images); \textbf{c,} fluorescent images (n=122 independent images); \textbf{d,} phase-contrast images (n=120 independent images); \textbf{e,} DIC images (n=60 independent images). The box plots display descriptive statistics with the median value represented by the horizontal line within the box, the lower and upper quartiles delineating the borders of the box, and the vertical black lines indicating the 1.5 interquartile range.
\textbf{f,} Example segmentation results of the four microscopy image modalities: brightfield, fluorescent, phase-contrast, and DIC images (from top to bottom). \textbf{g,} Quantitative comparison on the post-challenge testing set (N=157). The box plot shows descriptive statistics across the post-challenge testing cases, with the median value represented by the horizontal line within the box, the lower and upper quartiles delineating the borders of the box, and the vertical black lines indicating the 1.5 interquartile range.
Cellpose-pretrain: Cellpose pretrained model (``cyto2"). 
Cellpose-scratch: Cellpose model trained from scratch on the challenge dataset.
Cellpose-finetune: Cellpose fine-tuned model on the challenge dataset.
Omnipose-pretrain: Omnipose pretrained model (``cyto2").
Omnipose-scratch: Omnipose model trained from scratch on the challenge dataset.
Omnipose-finetune: Omnipose fine-tuned model on the challenge dataset. 
}\label{fig:pk}
\end{figure*}

\subsection*{The best-performing algorithms outperform state-of-the-art cell segmentation algorithms}
In order to demonstrate the advancement of the winning algorithm beyond the state-of-the-art (SOTA) in cell segmentation, 
we conducted a comparative analysis involving the top three algorithms from our challenge, the leading algorithm KIT-GE~\cite{KIT-GE} from the CTC cell segmentation task, and two widely recognized pre-trained generalist models, Cellpose~\cite{CellPose21-NM} and Omnipose~\cite{cutler2021omnipose}. 
The comparisons also included model variants of Cellpose and Omnipose that were trained from scratch on our challenge dataset. The aim was to determine if the performance improvement mainly resulted from the training set.
Recognizing the importance of transfer learning-based algorithms in achieving a universal solution, we further collected a new external testing set with 157 diverse yeast and bacteria cell images (Methods, Supplementary Table 4) to thoroughly compare the top three algorithms with the fine-tuned Cellpose (namely, Cellpose 2.0~\cite{stringer2022cellpose2}) and Omnipose models as well. For both Cellpose and Omnipose, we used the ``cyto2" model checkpoints, recognized for their exceptional generalizability, as the pretrained model and the foundation for further fine-tuning.

% We trained KIT-GE and Cellpose 2.0 on the challenge training set and then predicted the testing images. The Cellpose 2.0 model was initialized with the default pre-trained Cellpose model. The pre-trained generalist models in Cellpose and Omnipose were directly employed to segment the testing images (Methods). Notably, these algorithms share the commonality of being constructed on CNN architectures, such as U-Net~\cite{UNet-MICCAI15}. 

Fig.~\ref{fig:pk}a (Supplementary Table 6) illustrates the F1 scores of these eight methods on the testing set, revealing that the top three best-performing algorithms achieved significantly higher accuracy than the existing SOTA algorithms. 
Specifically, The T1 algorithms achieved a median F1 score of 89.7\% (IQR: 36.7-82.4\%), surpassing the KIT-GE, Cellpose-pretrain, Cellpose-scratch, Omnipose-pretrain, and Omnipose-scratch by 49.9\%, 24.4\%, 35.4\%, 58.9\% and 48.7\%, respectively.

Fig.~\ref{fig:pk}b presents the results on the brightfield images, where the top two best-performing algorithms remained at the forefront, achieving median F1 scores of 91.4\% (IQR: 88.0-94.9\%) and 91.0\% (IQR: 86.1-93.8\%), respectively.
The Cellpose-finetune model exhibited comparable performance to the third-best-performing algorithm, achieving a significant improvement of 16.6\% in median F1 score over the Cellpose-pretrain model, as anticipated due to its training on the challenge dataset. 
Fig.~\ref{fig:pk}c shows the results on the fluorescent images, where the third best-performing algorithm outperformed others with a median F1 score of 80.8\% (IQR: 71.6-91.5\%), followed by the best-performing algorithm and the Cellpose-pretrain model. However, the F1 score of Cellpose-scratch and Omnipose-scratch declined substantially by 44.1\% and 7.4\%, respectively. This decrease can be attributed to the testing images being from new cell types not present in the training set.

In Fig.~\ref{fig:pk}d, the results for PC images demonstrated that the top three best-performing algorithms maintained their superiority in this category, achieving median F1 scores of 93.6\% (IQR: 87.9-96.4\%), 88.8\% (77.7-96.4\%), and 90.3\% (84.3-95.0\%), respectively. The CTC challenge's top-performing segmentation algorithm, KIT-GE, excelled in PC images due to its design for label-free images and the relatively simple segmentation of round-shaped cells.
Fig.~\ref{fig:pk}e shows the results on DIC images with the top three best-performing algorithms once again achieving the highest performance, achieving median F1 scores of 86.8\% (IQR: 83.5-88.0\%), 75.0\% (IQR: 68.9-78.1\%), and 80.3\% (77.2-85.1\%), respectively. While Omnipose-scratch yielded the best performance among the SOTA methods with a median F1 score of 43.4\% (IQR: 33.1-60.9\%), it still fell significantly behind the top three best-performing methods. Conversely, KIT-GE and Cellpose struggled in this category, because the DIC testing images were from new biological experiments and exhibited very low contrast.

We further visualized segmentation examples of the seven algorithms to gain insights into their characteristics (Fig~\ref{fig:pk}f, Extended Data Fig. 1). The top three best-performing algorithms demonstrated relatively robust results, with the best-performing algorithm (T1-osilab) displaying exceptional accuracy across diverse microscope types, cell types, and image contrasts. Notably, KIT-GE exhibited better performance on phase-contrast images than stained images, as it was designed based on a label-free challenge dataset. Nevertheless, KIT-GE struggled to segment other images from new biological experiments, indicating limited generalization ability in this context. The Cellpose models outperformed Omnipose models on most images, except for DIC images featuring numerous small objects with low contrasts. Additionally, Cellpose-scratch model surpassed the Cellpose-pretrain on brightfield images, exhibiting fewer segmentation errors. However, its performance decreased on other modalities that contained previously unseen images, leading to an increased number of missed cells in the segmentation results.

Finally, we conducted a post-challenge analysis by evaluating the top three algorithms, KIT-GE, and three variants each of Cellpose and Omnipose models (pre-trained, trained from scratch, and fine-tuned by transfer learning) using a new testing set comprising unseen images (Methods). As shown in Fig.~\ref{fig:pk}g (Supplementary Table 7, Fig. 2), the top three algorithms outperformed others, achieving median F1 scores of 95.0\% (IQR: 93.0-97.9\%), 88.7\% (IQR: 72.1-93.9\%), and 93.3\% (IQR: 83.7-96.7\%), respectively. 
Notably, the fine-tuned Cellpose and Omnipose models surpassed their scratch-trained counterparts by 12.2\% and 11.7\%, respectively, demonstrating the value of previously learned features in new learning contexts. However, their performances were still lower than the original pre-trained models. This discrepancy is largely attributed to the testing images originating from new sources, leading to a case of catastrophic forgetting during the fine-tuning process, a common phenomenon in transfer learning~\cite{review-continualLearn19,review-continualLearn-PAMI}.

\section*{Discussion}
The primary and arguably most notable observation in this challenge is the unequivocal superiority of the Transformer-based algorithm, which exhibited significantly enhanced performance compared to existing SOTA cell segmentation algorithms.
Transformers offer several unique advantages compared to CNNs. First, Transformers~\cite{attention-nips17} use self-attention mechanisms that can capture global context and long-range dependencies in images, while CNNs usually process local image patches. Second, Transformers have a larger model capacity than CNNs~\cite{ViT2020}, enabling them to learn intricate patterns and model nuanced relationships in images, which are essential for accurate cell segmentation. Third, Transformers excel in transfer learning settings, allowing the model to pre-train on large datasets and subsequently fine-tuned on specific downstream tasks or new datasets with limited annotations. Notably, this effective strategy was also successfully adopted by the winning algorithm.

The winning algorithm demonstrated a remarkable level of superiority compared to the leading algorithm from the CTC challenge, even after the latter was retrained on our dataset. This notable improvement can be attributed to the unparalleled diversity of the dataset. Unlike the CTC challenge dataset, which only comprised label-free images, our challenge dataset encompassed both labeled and label-free images. Furthermore, our challenge focused on universal segmentation algorithms, while the top-performing teams of the CTC challenge developed tailored models for each dataset~\cite{CTC-Seg-1st,nnunet}. This fundamental difference in strategy likely contributed to the substantial performance gap.

In addition to the Transformer-based architecture, we also identified several useful strategies for achieving top performance. First, different from the typical detection-then-segmentation paradigm~\cite{MaskRCNN17}, multi-head outputs were employed by most of the top algorithms~\cite{top1,top2,top3}, which converted the instance segmentation task into distance map regression tasks and a cell foreground semantic segmentation task, followed by post-processing to merge the output as instance labels. This approach was conclusively demonstrated to be superior to the conventional detection-then-segmentation paradigm in this challenge.
Another crucial aspect was the adoption of diverse and robust data augmentation techniques, which is important to improve the model generalization ability and reduce overfitting. In addition to commonly used global intensity augmentations (e.g., scaling, noise addition, and blurring) and spatial augmentations (e.g., rotation, zooming, flipping), participants introduced innovative augmentation methods. For example, Lee et al.~\cite{top1} employed cell-wise random perturbations in image intensity, while Li et al.~\cite{top4} used Mosaic data augmentation~\cite{yolov4}, enabling the model to learn object identification at varying scales.
Moreover, employing efficient backbone networks, such as SegFormer~\cite{xie2021segformer} and ConvNext~\cite{ConvNext}, offered a favorable accuracy-efficiency trade-off. The winning algorithm also demonstrated that the slide-window-based method was an efficient strategy for scalable inference (Supplementary Table 8). Specifically, the input image was partitioned into multiple smaller patches, and their predictions were subsequently stitched together to form the final label map. This method proved particularly crucial for whole-slide image segmentation, considering the inherent limitations of RAM and GPU memory in real practice.

Additionally, all the top three teams explored the potential of leveraging the unlabeled images to improve the segmentation performance. Specifically, T1-osilab~\cite{top1} employed consistency regularization~\cite{consistency-loss} to match the algorithm’s predictions on the clean and degraded unlabeled images and introduced an additional head module to reconstruct the unlabeled images~\cite{reconstruction-loss}. Both T1-osilab~\cite{top1} and T2-sribdmed~\cite{top2} investigated pseudo-label learning, generating pseudo labels for unlabeled images using trained models, followed by training the network with both pseudo labels and ground-truth annotations. T3-cells~\cite{top3} implemented the uncertainty-aware Listen2Student mechanism~\cite{Listen2Student} to train a student network with low-uncertainty pseudo labels. However, despite these joint efforts, none of the employed methods demonstrated a notable enhancement in segmentation performance. Thus, it remains an open question how to effectively use unlabeled data to boost cell segmentation performance.

Furthermore, we made a noteworthy observation concerning the commonly employed transfer learning algorithm, which exhibited a phenomenon known as catastrophic forgetting~\cite{review-continualLearn-PAMI}. The original Cellpose and Omnipose generalist models, pre-trained on a diverse array of microscopy images, demonstrated the ability to generalize to a portion of the unseen testing images. However, their fine-tuned counterparts, exhibited a notable performance degradation, as they could only segment images present in the training set, while losing previously learned capability to handle unseen images.
The winning algorithm addressed this issue by implementing a simple yet effective strategy known as cell memory replay\cite{memory-replay}, aiming to re-learn the existing data during fine-tuning. More specifically, the fine-tuning procedure involved combining images from both the existing dataset and the new dataset as a mini-batch for training the model, allowing the algorithm to retain its competence in handling both known and novel images.

To promote the widespread applicability of the new SOTA algorithms, all top-performing teams have made their algorithms publicly available on GitHub, complete with comprehensive preprocessing, training, and testing code. However, a critical challenge remains in bridging the gap between these advanced algorithms and their seamless integration into daily biological practice, as it often demands a basic level of computational expertise to apply these algorithms to new images successfully.
To bridge this gap, we invited the top three best-performing teams to integrate their algorithms into Napari~\cite{napari}, an open-source interface specifically designed for user-friendly biological image visualization and analysis.
In this way, users gain convenient access to these high-performing algorithms, enabling them to effortlessly apply the segmentation techniques to their own images without necessitating additional coding. 
Furthermore, to facilitate even greater accessibility and ease of use, the algorithm Docker containers were thoughtfully released. This strategic move empowers users to perform batch image segmentation with utmost simplicity, as a single-line command suffices to initiate the process. 

These new cell segmentation algorithms have many potential applications in various biological tasks. For example, cell segmentation in mass cytometry imaging, as demonstrated in Jackson, Hartland W., et al.~\cite{Nature2020Breast}, was pivotal in characterizing cellular phenotypes in breast tumor tissues. These phenotypes, aligning closely with pathologist-assigned tumor grades, revealed complex multicellular structures. Similarly, cell segmentation played a crucial role in quantifying molecules at a single-cell level, as seen in the work of Capolupo, Laura, et al.~\cite{Science22Lip}, leading to the discovery of novel regulatory mechanisms in dermal fibroblasts. Additionally, the application of cell segmentation in disease progression studies, such as those by Risom, Tyler et al.~\cite{cell-breastcancer}, has been instrumental in characterizing cancer microenvironments in MIBO-TOF imaging of tissue microarrays.

This work has certain limitations.  
While the challenge dataset was indeed diverse, it was confined to 2D microscopy images due to the available datasets. However, three-dimensional (3D) microscopy images are becoming increasingly prevalent~\cite{NMeth3D-Img}, which pose new segmentation challenges, such as the large-scale volume and anisotropic resolutions. Additionally, while the integration of Napari cell segmentation interfaces has improved accessibility for biologists, these algorithms currently do not support interactive feedback from users. Furthermore, the scope of the challenge was restricted to segmentation tasks, omitting classification tasks.
Future endeavors should aim to broaden the benchmark to include more complex 3D images, coupled with classification tasks. There is also a compelling need to develop a biologist-in-the-loop system, enabling more effective collaboration between algorithms and human experts.

In conclusion, the challenge results present a successful proof of concept of generalist cell segmentation algorithms, benefiting from the collective expertise of both biological imaging and machine learning experts. The Transformer-based algorithm surpassed previous SOTA methods by a large margin, which can efficiently generate accurate cell contours on a wide range of microscopy images without user intervention. 
Furthermore, the top algorithms have been made open-source and seamlessly integrated into user-friendly interfaces. This integration holds great potential for accelerating microscopy image analysis throughput and fostering new discoveries in quantitative biological research.
We aim to establish this challenge as a sustainable benchmark platform and we enthusiastically welcome contributions of various new data to expand the data diversity, paving the way for continuous advancement in this vital field.

\subsection*{Acknowledgements} 
This work was supported by the Natural Sciences and Engineering Research Council of Canada (NSERC, RGPIN-2020-06189 and DGECR-2020-00294) and CIFAR AI Chair programs. This research was enabled in part by computing resources provided by the Digital Research Alliance of Canada.
We thank Patrick Byrne, Maria Kost-Alimova, Shantanu Singh, and Anne E. Carpenter for contributing U2OS and adipocyte images.
We thank Dr. Andrea J. Radtke and Dr. Ronald Germain for contributing adenoid and tonsil whole-slide fluorescent images.
We thank Sweta Banerjee for providing multiple myeloma plasma cell annotations in stained brightfield images. 
The platelet differential interference contrast images collected by Carly Kempster and Alice Pollitt were supported by the British Heart Foundation/NC3Rs (NC/S001441/1) Grant. Anubha Gupta would like to thank the Department of Science and Technology, Govt. of India for the SERB-POWER fellowship (Grant No.: SPF/2021/000209) and the Infosys Centre for AI, IIIT-Delhi for the financial support to run this challenge.
ML, VG, MS, SJR were supported by SNSF grants CRSK-3\_190526 and 310030\_204938 awarded to SJR.
E. Upschulte and T. Dickscheid received funding from Priority Program 2041 (SPP 2041) ``Computational Connectomics" of the German Research Foundation (DFG), and the Helmholtz Association’s Initiative and Networking Fund through the Helmholtz International BigBrain Analytics and Learning Laboratory (HIBALL) under the Helmholtz International Lab grant agreement InterLabs-0015. The authors gratefully acknowledge the computing time granted through JARA on the supercomputer JURECA at Forschungszentrum J\"ulich.
We also thank the grand-challenge platform for hosting the competition. 

\subsection*{Author Contributions Statement} 
J.M. conceived and designed the analysis, collected and cleaned the data, contributed analysis tools, managed challenge registration and evaluation, performed the analysis, wrote the initial manuscript, and revised the manuscript; R.X. conceived and designed the analysis, managed challenge registration and evaluation, and revised the manuscript; S.A., C.G., A.G., R.G., S.G., and Y.Z. conceived and designed the analysis, cleaned data, contributed labeled images, managed challenge registration and evaluation, performed the analysis, and revised the manuscript. 
G.L., J.K., W.L., H.L., E.U., and T.D. participated in the challenge, developed the top-three algorithms, and made the code public available;
J.G.A., Y.W., L.H., X.Y. cleaned data, contributed labeled images, and managed challenge registration and evaluation; 
M.L., V.G., M.S., S.J.R., C.K., A.P., L.E., T.M. J.M.M., J.-N.E., contributed new labeled data the the competition. W.L., Z.L., X.C., and B.B. participated in the challenge, developed algorithms, and made the code public available;
N.F.G., D.V.V., E.W., B.A.C., and O.B contributed public or unlabeled data to the competition. T.C. managed challenge registration and evaluation.
G.D.B. and B.W. conceived and designed the analysis, wrote and revised the manuscript.

\subsection*{Competing Interests Statement} 
Song Gu is employed by Nanjing Anke Medical Technology Co., Ltd. 
Jan Moritz Middeke and Jan-Niklas Eckardt are co-owner of Cancilico. 
David Van Valen is a co-founder and Chief Scientist of Barrier Biosciences and holds equity in the company. 
Oscar Br{\"u}ck declares the following competing financial interests: consultancy fees from Novartis, Sanofi, and Amgen, outside the submitted work; research grants from Pfizer and Gilead Sciences, outside the submitted work; stock ownership (Hematoscope Ltd), outside the submitted work.
All other authors have no competing interests.

\newpage
\section*{Methods}
\subsection*{Challenge Organization}
This challenge was pre-registered in the Thirty-sixth Conference on Neural Information Processing Systems (NeurIPS) (\url{https://neurips.cc/Conferences/2022/CompetitionTrack})  following a peer review process conducted by the competition program committee comprising experts in both machine learning and challenges organization. The challenge was officially launched on June 15, 2022, and ran for 139 days until October 31, 2022, marking the testing submission deadline. 
Throughout the development phase, participants were given the opportunity to submit their tuning set segmentation results on the challenge platform and obtain corresponding F1 scores. Moreover, to minimize entry barriers, we supplied a U-Net-based baseline model and offered a step-by-step tutorial to assist participants in becoming familiar with model training, inference, and submission. Additionally, we furnished guidelines and suggestions on state-of-the-art cell segmentation methods~\cite{CellPose21-NM,cutler2021omnipose,stringer2022cellpose2,Mesmer21-NatBio}, empowering participants to surpass the baseline and achieve higher levels of performance.

\subsection*{Dataset curation and pre-processing}
The images were curated from multiple laboratories, each specializing in different cell types and modalities. Five labeled datasets were obtained from publicly available datasets with proper license permits or approval from the respective authors. Two public datasets contained a subset of labeled images and we augmented them with complete cell annotations for the previously unlabeled images. The annotation process was conducted in the original data center by our author team. 
Eight public datasets lacked annotations and we generated cell annotations for the challenge. All testing images were newly acquired by ourselves for this challenge. 
It should be noted that the testing images remained unavailable to the public during the challenge, avoiding potential data or label leakage. 
In addition to the primary testing set, we further collected a new batch of yeast (e.g., \emph{clb1-6}$\Delta$ conditional mutants and pseudohyphal cells) and bacteria images (e.g., Myxoccocus xanthus and Ruminiclostridium cellulolyticum) for post-challenge analysis, aiming to evaluate the algorithms' accuracy and robustness when applied to unseen cells exhibiting a diverse range of appearances and morphologies.
Detailed sources can be found in Supplementary Table 1-4. We also presented the distribution of image size in Extended Data Fig. 3. The majority of the images are microscope patches, but several whole-slide images are also provided in each set. The training (labeled) set, tuning set, and testing set exhibit a similar median image size, approximately one million pixels (1000x1000).

The original image formats included png, bmp, jpg, tif, tiff, npy, and npz. The npy and npz formats, which are not typical image formats, were converted to the widely used png format. All other image formats were retained as they were, to accommodate the diverse array of image formats that the developed algorithms might encounter.
External datasets and pre-trained models are allowed, but participants should post the corresponding links to the competition forum and we also maintained a document of external datasets on the challenge homepage to make sure these external datasets were available to all participants.

For the labeled dataset, all the cells were annotated in each image, with the exception of red blood cells in blood and bone marrow slides since biomedical researchers predominantly centered on the stained leukocytes. 
The annotation team consisted of two biologists with 10 years of experience, responsible for ensuring compliance with annotation requirements. In cases where data contributors provided cell annotations, the annotations were thoroughly checked and revised as needed. For contributors who provided unlabeled images, publicly available specialist models~\cite{Yeaz, YeastMate, Mesmer21-NatBio, DIC-SciRep} were initially employed to generate predictions. The resulting segmentation outcomes were subsequently subjected to manual revision by the biologists. Additionally, to maintain the quality and reliability of the dataset, each image-annotation pair underwent stringent quality control. Images with less than five cells were excluded from the dataset, and cells containing fewer than 15 pixels were also removed. In total, we created more than 900,000 new cell annotations for the challenge, which is significantly larger than the provided new annotations in the recent CoNIC challenge~\cite{CoNIC} and cell tracking challenge~\cite{celltrack-nm23}.

\subsection*{Top three best-performing algorithms}
\textbf{Best-performing algorithm.}
Lee et al.~\cite{top1} (T1-osilab) incorporated model-centric and data-centric approaches to learn generalizable representations for heterogeneous microscopy image modalities and achieved a good trade-off between model accuracy and efficiency. 
From the model-centric perspective, the framework adopted a typical encoder-decoder architecture to extract hierarchical features and integrate them through skip connections. Concretely, SegFormer~\cite{xie2021segformer} served as the encoder, while MA-Net~\cite{MANet} was employed as the decoder, utilizing the Mish~\cite{misra2020mish} activation function. The network jointly predicted cell probability maps and regressed cell-wise vertical and horizontal gradient flows, followed by a gradient tracking post-processing to separate touched cells, which was originally proposed in Cellpose~\cite{CellPose21-NM}. 
From the data-centric perspective, they tailored two cell-aware augmentations to extensively enrich the diversity of the dataset and combined them with commonly used intensity and spatial augmentation methods to improve model generalization. Specifically, image intensities were randomized in a cell-wise manner and cell boundary pixels were excluded to separate the crowded cells. Moreover, a two-phase pre-training and fine-tuning pipeline was used to retrain the knowledge from external datasets, including TissueNet~\cite{Mesmer21-NatBio}, Omnipose~\cite{cutler2021omnipose}, Cellpose~\cite{CellPose21-NM}, and LiveCell~\cite{edlund2021livecell}. Furthermore, to address minor modalities, they were selected through unsupervised clustering with the latent embedding and subsequently over-sampled during training, thereby aiming to enhance the performance of these less-represented modalities. 

The model inputs were three-channel images. The overall loss function was the combination of binary cross-entropy loss and mean-square error loss. The inference process relied on the sliding window strategy, a highly efficient approach for processing whole-slide images. During the merging of predictions from these small window patches, an importance map was generated and applied to the predictions, thereby preventing the recognition of the same cells at the patch boundary as multiple cells. 
The comprehensive integration of these approaches resulted in exceptional performance, effectively handling diverse microscopy image modalities with high accuracy.

\textbf{Second-best-performing algorithm.}
Lou et al. ~\cite{top2} (T2-sribdmed) designed a classification-and-sengmentation framework that first classified the input image into one group and then performed cell segmentation with a model trained for that group. The classification pipeline consisted of three steps. First, it employed a segmentation model trained on labeled images to generate pseudo labels for unlabeled images. 
Second, the images were classified into four groups based on image intensities. Specifically, the first class included all single-channel images. The three-channel RGB images were converted to Hue, Saturation, and Value (HSV) color space. Within this transformed domain, images exhibiting a mean saturation (S) greater than $0.1$ and a mean value (V) falling within the range $[0.1, 0.6]$ were assigned to the second class. 
The remaining images with cell areas larger than $8000$ pixels were classified as the third class, while others were designated as the fourth class. 
Finally, a ResNet18~\cite{ResNet16} was trained for automated group classification.
The segmentation network followed a design of U-Net-like architecture, where the encoder was ConvNeXt~\cite{ConvNext}. Motivated by the observation that most cells in the first and second classes were roundish, a decoder with star-convex polygon-based cell representation~\cite{Stardist} was integrated with the encoder for cell instance segmentation, termed as ConvNeXt-Stardist. The non-maximum suppression (NMS)~\cite{FasterRCNN} was employed in the post-processing to remove duplicated predictions. 
For the third and fourth classes, the prediction head in HoverNet~\cite{CoNSeP} was adopted as the decoder, termed as ConvNeXt-Hover. The marker-based watershed algorithm was applied in the post-processing phase to separate touching cells.

There were four segmentation models in total trained for four image groups respectively. ConvNeXt-Stardist was trained with a combination of cross-entropy loss, Dice loss, and MAE loss, and ConvNeXt-Hover was trained with a combination of cross-entropy loss, Dice loss, MSE loss, and mean squared gradient error loss. Both ConvNeXt-Stardist and ConvNeXt-Hover were pre-trained on all images and finetuned on the images from the corresponding group. The model inputs were three-channel images. During inference, the input image was first classified into certain groups by the classification model and then processed by the segmentation model trained for the group.

\textbf{Third-best-performing algorithm.}
Upschulte et al.~\cite{top3} (T3-cells) proposed a Contour Proposal Network (CPN)~\cite{CPN}, which treated instance segmentation as a sparse detection problem by regressing object contours anchored at pixel locations. This enabled the model to handle multiple objects assigned to the same pixel and recover partially superimposed objects accurately. The shape-focused nature of contour representation learning also facilitated the development of inductive shape priors, potentially improving robustness in challenging conditions.

The CPN utilized the ResNeXt backbone network~\cite{ResNeXt} to extract multiscale feature maps, a regression head to generate candidate contour representations for each pixel, and a classification head to determine whether an object was present or not at these locations. A proposal sampling stage extracted a sparse list of contour representations, which were transformed into the pixel domain using differentiable Fourier transformation to encode contour information in the frequency domain~\cite{FourierTransform}. The precision of the contours was further improved by using a displacement field generated by an additional regression head. In addition to the original CPN, this work introduced dedicated supervision for boundaries and proposed an extra branch to estimate localization uncertainty for boundaries. The multitask training objective was defined by a combination of the average absolute difference loss for contour regression, the generalized IoU loss for boundary localization~\cite{GIoU}, the absolute L1 distance for local refinement~\cite{CPN}, the distance loss for frequency regularization~\cite{CPN}, the binary cross entropy loss for classification, and the negative power log-likelihood loss for uncertainty estimation~\cite{NPLL}.

The uncertainty-aware Listen2Student mechanism~\cite{Listen2Student} was applied to incorporate unlabeled examples during training, where a teacher model generated bounding boxes as pseudo-labels to supervise the student model. The model inputs were three-channel images. For post-processing, the Vanilla NMS relying solely on the classification score might not reliably indicate the proposal's quality. To address this issue, the approach proposed in~\cite{NPLL} was employed to incorporate uncertainty estimations into the NMS selection process. The object contours were transformed into segmentation masks through rasterization and region filling. A region-growing technique~\cite{adams1994seeded} was further adopted for overlapping regions.

\subsection*{Existing SOTA cell segmentation algorithms}
The following methods were designed for grey and two-channel microscopy images, while the challenge dataset was curated for developing universal algorithms that were agnostic to different image channel formats. Thus, we preprocess the challenge images to grey images by the `skimage.color.rgb2gray` function.  

\textbf{Cellpose}~\cite{CellPose21-NM} represents a significant advancement in the field of general cellular segmentation algorithms. It used U-Net~\cite{UNet-MICCAI15} to predict horizontal and vertical gradient maps of cell instances and a foreground binary mask. After that, individual cells are segmented by grouping the pixels that point to the same center point in the gradient maps. This unique design allows it to be capable of processing a wide variety of cell morphologies in a unified framework.
In the comparative studies, we used the most generalizable ``cyto2" model as the pre-trained model, which was trained on Cellpose dataset and user-submitted images.

\textbf{Omnipose}~\cite{cutler2021omnipose} was an extension of Cellpose, aiming to handle very elongated cells, especially bacterial cells. The network architecture backbone was still U-Net but the model had four heads to predict four components: two gradient flows, a distance transform map, and a boundary map. 
We also chose the ``cyto2" model to infer the challenge testing images. 

\textbf{Cellpose 2.0}~\cite{stringer2022cellpose2} further introduced a transfer learning-based method, an important branch towards general cell segmentation solutions, to quickly adapt the pre-trained models to new microscopy images. With a human-in-the-loop pipeline, users can train customized cellular segmentation models by fine-tuning pre-trained Cellpose models with only 100-200 annotated regions of interest. The network architecture in Cellpose 2.0 was the same as the Cellpose model.

\textbf{KIT-GE}~\cite{KIT-GE} trained a U-Net model to predict cell distance and neighbor distance, followed by watershed post-processing. Compared to the original U-Net~\cite{UNet-MICCAI15}, the maximum pooling layers were replaced with 2D convolutional layers with stride 2, and batch normalization layers were added after the convolutional layers.

\subsection*{Evaluation metrics}
This challenge focused on two key metrics: segmentation accuracy and efficiency. While segmentation accuracy is a fundamental metric in cell segmentation, we included efficiency in the evaluation to account for its significance during model deployment. If the challenge metrics only considered the algorithm accuracy, participants may solely prioritize it by employing the ensemble of multiple models~\cite{NM-DSB18}. However, such solutions may not be practical in real-world scenarios, particularly for biologists who typically have limited computational resources. Recognizing this, we incorporated efficiency as an evaluation metric to guide participants in considering the trade-off between model accuracy and efficiency.

\textbf{Segmentation accuracy metric: F1 score.}
Cell segmentation is a typical instance segmentation task. We employed the widely used F1 Score to evaluate the segmentation results~\cite{Mesmer21-NatBio, maier2022metrics, NM-SegMetric}. Specifically, each predicted cell mask is matched to the most similar ground-truth mask based on the predefined intersection over union (IoU) threshold (i.e., 0.5). A predicted cell mask is classified as correct segmentation as long as its IoU is over the predefined IoU threshold. A higher threshold requires a larger overlap between the predicted cell mask and the ground-truth mask, and a commonly used threshold is 0.5. Then, all the cells can be divided into three categories, including true positives (TP), false positives (FP), and false negatives (FN). TP denotes correctly segmented cells. FP denotes wrong segmented cells while FN denotes missed cells in the segmentation mask. 
After that, we can compute the precision and recall, which are defined by $precision = \frac{TP}{TP+FP}$ and $recall = \frac{TP}{TP+FN}$, respectively. 
The F1 score can be interpreted as a harmonic mean of the precision and recall, which is defined by 
\begin{equation*}
    F1 = \frac{2\times Precision \times recall}{precision + recall}.
\end{equation*}
Since the cells located in the boundaries are usually incomplete and have low practical value in various downstream analysis tasks, we remove these cells when computing the metrics.

\textbf{Segmentation efficiency metric: Running time.}
All the submitted Docker containers were run on the same desktop workstation with a 12-core CPU, 32GB RAM, and one NVIDIA 2080Ti GPU.
In order to obtain the running time $T$ for each image, the testing images were segmented one by one. To compensate for the Docker container startup time, we gave a time tolerance for the running time. Specifically, the time tolerance was 10s if the image size (height H $\times$ width W) was no more than 1,000,000. If the image size was more than 1,000,000, the time tolerance was $(H\times W) / 1000000 \times 10s$. This time tolerance was determined by the open-source U-Net baseline.

\subsection*{Ranking Scheme: Rank-then-aggregate}
Both F1 score and running time will be used for ranking. 
However, the two metrics cannot be directly fused because they have different dimensions. Thus, we will use a ``rank-then-aggregate" scheme for ranking, including the following three steps:
\begin{itemize}
    \item Step 1. Computing the two metrics for each testing case and each team;
    \item Step 2. Ranking teams for each of the $N$ testing cases such that each team obtains N$\times$2 rankings;
    \item Step 3. Computing ranking scores for all teams by averaging all these rankings and then normalizing them by the number of teams. The final rank will be determined by the mean ranking scores.
\end{itemize}

In addition to the employed rank-then-aggregate scheme, several other strategies can be used to obtain a ranking, but these may lead to different orderings of algorithms and thus different winners~\cite{lena-2018rankings}. 
A typical ranking scheme was ``aggregate-then-rank": computing mean scores across all testing cases for each team and then using this aggregation to rank each team.
One can also use test-based procedures for ranking. Specifically, each pair of algorithms are compared by statistical hypothesis tests. The ranking is then performed according to the resulting relations or according to the number of significant one-sided test results. In the latter case, if algorithms have the same number of significant test results, then they obtain the same rank. For analysis purposes,  we computed the ranks of the 28 algorithms based on five different ranking
schemes: mean-then-rank, median-then-rank, rank-then-mean, rank-then-median, and statistical significance test-based ranking.

Importantly, for a transparent challenge, the evaluation code and ranking scheme were publicly available at the beginning of the challenge. For comparative analysis, we applied different ranking schemes to the 28 algorithms, including rank-then-mean, rank-then-median, median-then-rank, mean-then-rank, and test-based rank. It can be found that most algorithms had fluctuations under different ranking schemes but the winning algorithm consistently held the first place.

\subsection*{Ranking Stability and statistical analysis}
Ranking stability is an important factor for robust challenge results~\cite{challengeR}. Thus, we applied bootstrapping and computed Kendall’s $\tau$~\cite{kendall-tal} to quantitatively analyze the variability of our ranking scheme. Specifically, we first extracted 1000 bootstrap samples from the international validation set and computed the ranks again for each bootstrap sample. Then, the ranking agreement was quantified by Kendall’s $\tau$. 
Kendall’s $\tau$ computes the number of pairwise concordances and discordances between ranking lists. Its value ranges $[-1, 1]$ where -1 and 1 denote inverted and identical order, respectively. A stable ranking scheme should have a high Kendall’s $\tau$ value that is close to 1.
To compare the performance of different algorithms, we performed Wilcoxon signed rank test because it is a paired comparison. Results were considered statistically significant if the $p-value$ is less than 0.05.

\section*{Data availability}
The dataset has been publicly available on the challenge website \url{https://neurips22-cellseg.grand-challenge.org/}. It is also available at the Zenodo repository~\url{https://zenodo.org/records/10719375}.

\section*{Code availability}
The top ten teams have made their code publicly available at GitHub \url{https://neurips22-cellseg.grand-challenge.org/awards/}. The code of evaluation metrics and existing cell segmentation algorithms are available at \url{https://github.com/JunMa11/NeurIPS-CellSeg}.
The Napari interface and huggingface online segmentation tool for the top three best-performing algorithms are available at
T1-osilab:~\url{https://github.com/joonkeekim/mediar-napari};
T2-sribdmed:~\url{https://github.com/Lewislou/cellseg_sribd_napari}; 
T3-cells:~\url{https://github.com/FZJ-INM1-BDA/celldetection-napari}.
They are also available at the Zenodo repository~\url{https://zenodo.org/records/10718351}.

\newpage

\renewcommand{\refname}{Methods-only references}

\end{document}